\documentstyle[12pt]{article}

\begin{document}
\baselineskip = 24pt

\begin{titlepage}

\hoffset = .5truecm
\voffset = -2truecm

\centering

\null
\vskip -1truecm
\rightline{IC/94/401}
\vskip 1truecm

{\normalsize \sf \bf International Atomic Energy Agency\\
and\\
United Nations Educational, Scientific and Cultural Organization\\}
\vskip 1truecm
{\huge \bf
INTERNATIONAL CENTRE\\
FOR\\
THEORETICAL PHYSICS\\}
\vskip 3truecm

{\LARGE \bf
Gravitational Lorentz anomaly\\ from the overlap formula in 2-dimensions
\\}
\vskip 1truecm

{\large \bf
S. Randjbar-Daemi
\\}
\medskip
{\large      and\\}
\medskip
{\large \bf
J. Strathdee
\\}

\vskip 6truecm

{\bf International Centre for Theoretical Physics \\}
December 1994
\end{titlepage}

\hoffset = -1truecm
\voffset = -2truecm

\title{\bf
Gravitational Lorentz anomaly\\ from the overlap formula in 2-dimensions
\\}
\vspace{2cm}

\author{
{\bf
S. Randjbar-Daemi}\\
\normalsize International Centre for Theoretical Physics, Trieste 34100,
{\bf Italy}\\
{\normalsize and}\\
\normalsize
{\bf}\\
{\bf J. Strathdee}\\
\normalsize International Centre for Theoretical Physics, Trieste 34100,
{\bf Italy} }

\date{December 1994}
\newpage

\maketitle

\begin{abstract}
In this letter we show that the overlap formulation of chiral gauge theories
correctly reproduces
the gravitational Lorentz anomaly in 2-dimensions. This formulation has been
recently suggested as a
solution to the fermion doubling problem on the lattice.The well known
response to general coordinate transformations of the effective action of
Weyl fermions coupled to gravity in 2-dimensions  can also be recovered.
\end{abstract}

\newpage


The formulation of lattice chiral gauge theories has been an outstanding
problem
for many years\cite{kn:Karsten.}. In a recent paper a new formulation of this
problem has
been suggested\cite{kn:Nar1.}. One of the necessary steps in checking the
viability of this
suggestion is to show that it correctly reproduces the chiral anomalies in the
continuum limit.
For the case of a $U(1)$ gauge theory in 2-dimensions this was done in
\cite{kn:Nar2.}, and its
generalization to non-abelian chiral anomalies in 4-dimensions is contained in
\cite{kn:Ran.}.

In this letter we would like to use the notation and formalism developed in
\cite{kn:Ran.} to
examine the anomalous coupling of Weyl fermions to a background gravitational
field in 2 dimensions.
It will be shown that the overlap formalism proposed in \cite{kn:Nar1.}
correctly reproduces the
chiral anomaly in this system.

Our starting point will be the Hamiltonian of a 2-component "massive" fermion
coupled to a
 gravitational field $e_{a}^ {\mu}$ in $2+1$ dimensions
\begin{equation}
H=\int d^{2}x\ \psi^{\dagger}(x)\sigma_{3}(\sigma_{a}e_{a}^ {\mu}\nabla_{\mu}
+\Lambda)\psi(x)
\label{eq:H}
\end{equation}
where $\sigma_{a}$, $a=1,2$ and $\sigma_{3}$ are the Pauli spin matrices and
$\nabla_{\mu}= \partial_{\mu}-{i\over 2} \omega_{\mu} -{1\over
2}\partial_{\mu}ln\ e$,
with $\omega_{\mu}$ being the spin connection derived from the zweibein $e_{a}^
{\mu}$ and
$e^{-1}=dete_{a}^ {\mu}$. All the fields in (\ref{eq:H}) depend only on the
2-dimensional
 coordinates.\\
The Hamiltonian (\ref{eq:H}) is invariant under general coordinate
transformations
$x^{\mu}\rightarrow f^{\mu}(x^{1}, x^{2})$ provided the 2-component spinor
field $\psi$ transforms
as a scalar density of weight ${1\over 2}$. It is also invariant under the
local
$O(2)$ frame rotations,
$\psi(x)\rightarrow e^{{i\over 2}\theta(x)\sigma_{3}}\psi(x)$.

It was shown in \cite{kn:Ran.} that in the limit of $|\Lambda|\rightarrow
\infty$ the
Hamiltonians of the type (\ref{eq:H}) can be used to recover the Green's
functions
of a massless chiral gauge theory. It is well known that the gravitational
coupling of
such fermions is anomalous\cite{kn:Alv.}. Here we would like to rederive this
anomaly
in the local frame rotations as $|\Lambda|\rightarrow \infty$.

The overlap formulation of \cite{kn:Nar1.} defines the effective action
$\Gamma[e]$ by
$$
\Gamma[e]= -ln <e;+|e;->
$$
where $|e;+>$ and $|e;->$ are the Dirac ground states of the two Hamiltonians
$H(+\Lambda)$  and
$H(-\Lambda)$, respectively. To study the behaviour of $\Gamma[e]$ under local
frame rotations we must test the response of the two ground states with respect
to
such transformations. Acting on the Schr\"odinger picture fields these
transformations
are realized by unitary operators $U(\theta)$,
$$
U(\theta)^{-1}\psi(x)U(\theta) = e^{{i\over 2}\theta(x)\sigma_{3}}\psi(x)
$$
Since the free Hamiltonian
$H_{0}=\int d^{2}x\ \psi^{\dagger}(x)\sigma_{3}(\sigma_{\mu}\partial_{\mu}
+\Lambda)\psi(x)$
is invariant under the constant gauge transformations it follows that
$$
U(\theta)^{-1}H(e)U(\theta) = H(e^{\theta})
$$
where $e^{\theta}$ denotes the rotated frame. We shall regard  $|e;\pm>$  as
the
perturbed vacuua of the Dirac ground states $|\pm>$  of $H_{0}(\pm\Lambda)$.
 For weak fields it then follows
$$
U(\theta)|e;\pm>=|e^{\theta};\pm>e^{i\Phi_{\pm}(\theta;e)}
$$
where the angles $\Phi_{\pm}(\theta;e)$  are real. It is possible to compute
these angles
by applying "time" independent perturbation theory. To first order in $\theta$
this gives
\cite{kn:Ran.}
$$
\Phi_{\pm}(\theta;e)=
\int d^{2}x\ <\pm|\psi^{\dagger}(x){1\over 2}\sigma_{3}
\theta (x)\psi(x)[1- {1\over {E_{0\pm}-H_{0\pm}}}\Pi_{\pm}(V-\Delta
E_{\pm})]^{-1}|\pm>
$$
where $\Pi_{\pm}=1-|\pm><\pm|$ and
$$
V=\int d^{2}x\ \psi^{\dagger}(x)\sigma_{3}[\sigma_{\mu}
(-{i\over 2} \omega_{\mu}+{1\over 2}\partial_{\mu}h)+\sigma_{a}
h_{a}^{\mu}\partial_{\mu}]\psi(x)
$$
Here the weak gravitational perturbations around the flat space are given by
$ h_{a}^{\mu}= e_{a}^{\mu}-\delta_{a}^{\mu}$ and $h=
\delta^{a}_{\mu} h_{a}^{\mu}$. The actual values of the vacuum shifts $\Delta
E_{\pm}$ are
irrelevant for the present discussion. For further detail of the notation
reference \cite{kn:Ran.}
can be consulted.
Since $U(\theta)$ is unitary it follows that the overlap must satisfy
$$
<e^{\theta};+|e^{\theta};->=<e;+|e;->e^{i(\Phi_{+}-\Phi_{-})}
$$
which implies an anomaly if $\Phi_{+}-\Phi_{-}\not= 0$. We  compute
this difference perturbatively and show that in the limit of
$|\Lambda|\rightarrow \infty$ it
contains the usual Lorentz anomaly.

The first order part of $\Phi_{+}$ is given by
\begin{equation}
\Phi_{+}^{(1)}=\sum_{n}\int_{\Omega} d^{2}x\ <+|\psi^{\dagger}(x){1\over
2}\sigma_{3}
\theta(x)\psi(x)|n>{1\over {E_{0+}-E_{n}}}<n|V|+>
\label{eq:fi}
\end{equation}
where the sums are restricted to $2$-particle intermediate states,
$$
\sum_{n} |n><n|={1\over \Omega
^{2}}\sum_{k_{1},k_{2}}|k_{1},k_{2}><k_{1},k_{2}|
$$
 and where
$$ |k_{1},k_{2}>= b_{+}^{\dagger}(k_{1})d_{+}^{\dagger}(k_{2})|+>
$$
The operators $b_{\pm}$ and $d_{\pm}$ are defined by \cite{kn:Ran.}
$$
\psi(x)={1\over \Omega}\sum_{k}(b_{\pm}u_{\pm}(k)
+d_{\pm}^{\dagger}v_{\pm}(k))e^{ikx}
$$
where $\Omega$ is the volume of a $2$-box and $u$'s and $v$'s are the positive
and the negative
energy eigenvectors of $H_{0\pm}(k)=
\sigma_{3}(i\sigma_{\mu}k_{\mu} \pm |\Lambda|)$,
 $$
H_{0\pm}(k)u_{\pm}(k)=\omega (k)u_{\pm}(k),\ \ \ \ \ \
H_{0\pm}(k)v_{\pm}(k)=-\omega (k)v_{\pm}(k)
$$
where $\omega(k)=(k^{2}+\Lambda^2)^{1\over 2}$.\\
By making use of these results in (\ref{eq:fi}) we obtain
\begin{eqnarray*}
{\delta\Phi_{+}^{(1)}\over \delta \theta(x)} &=&{i\over 4\Omega^{2}}
\sum_{k_{1},k_{2}}
{e^{i(k_{1}+k_{2})x}\over {\omega (k_{1})+\omega (k_{2})}}Tr\Bigl[
\sigma_{3}U(k_{1})\sigma_{3}\sigma^{\mu}
\{\tilde \omega (k_{1}-k_{2})\sigma_{3}+\Bigr.\\
&&\Bigl. (k_{1}-k_{2})_{\mu}\tilde h(k_{1}-k_{2}) -2k_{2\nu}\tilde
h_{\mu}^{\nu}((k_{1}-k_{2})
\}V(k_{2})
\Bigr]
\end{eqnarray*}
where $U(k)={\displaystyle\omega (k)+\sigma_{3}(i\sigma_{\mu}k_{\mu}
+\Lambda)\over
\displaystyle 2\omega (k)}= 1- V(k)$ and
the Fourier transforms are defined in the usual way, e.g.
$\tilde h(k)=\int_{\Omega} d^{2}xe^{-ikx}h(x)$
To obtain $\Phi_{-}^{(1)}$ we need to change the sign of $\Lambda$. Evaluating
the Dirac traces and
letting $\Omega\rightarrow\infty$ we obtain
\begin{equation}
{\delta(\Phi_{+}^{(1)}-\Phi_{-}^{(1)})\over \delta \theta(x)}=
\Lambda \int{d^{2}p\over 2(\pi)^{2}}\ e^{ikx}F_{\mu\nu}(p)\tilde h_{\nu\mu}(p)
\label{eq:an1}
\end{equation}
where
\begin{equation}
F_{\mu\nu}(p)=\int{d^{2}k\over 2(\pi)^{2}}
{\displaystyle k_{\mu}k_{\nu}\over \displaystyle {\omega ({k+p\over 2})}
\omega ({k-p\over 2})
\left(\omega ({k+p\over 2})+\omega ({k-p\over 2})\right)}
\label{eq:an2}
\end{equation}
Equation (\ref{eq:an1}) should yield the gravitational Lorentz anomaly. To
evaluate it we expand
$F_{\mu\nu}(p)$ in powers of ${1\over |\Lambda|}$ and obtain
\begin{equation}
F_{\mu\nu}(p)={|\Lambda|\over 2}\int{d^{2}q\over 2(\pi)^{2}}
{\displaystyle q_{\mu}q_{\nu}\over\displaystyle(q^{2}+1)^{3/2}}[1-
{3\over 8}{p^{2}\over \Lambda^2 ((q^{2}+1)} +{5\over8} {(q.p)^2\over
{\Lambda^2(q^{2}+1)^2}} +...]
\label{eq:an3}
\end{equation}
where all the terms not indicated explicitly are given by convergent integrals
and vanish as
$|\Lambda|\rightarrow\infty$. The leading term inside the bracket on the
right-hand side
makes a divergent contribution which must be subtracted in the usual way by
a suitable counterterm. It should be noted
that this contribution is independent of $p$ and therefore will not contribute
to the
terms involving the derivatives of $h_{\mu\nu}$.
Those of the convergent integrals which contribute in the limit of
$\Lambda\rightarrow\infty$
produce
\begin{equation}
F_{\mu\nu}(p)={1\over 48\pi|\Lambda|}
(p_{\mu}p_{\nu}-p^2\delta_{\mu\nu})
\label{eq:an4}
\end{equation}
Upon substitution of this result in (\ref{eq:an1}) we obtain
\begin{equation}
{\delta(\Phi_{+}^{(1)}-\Phi_{-}^{(1)})\over \delta \theta(x)}=
{\Lambda\over|\Lambda|} {1\over 48\pi}
(\partial^2\delta_{\mu\nu}-\partial_{\mu}\partial_{\nu}) h_{\nu\mu}(x)
\label{eq:an5}
\end{equation}
To see that this is the standard result we only need to make use of the
geometric relations
$\omega_{\mu}=-{1\over 2}e_{\mu}^a {\displaystyle\varepsilon^{\alpha\beta}\over
\displaystyle e}\partial_{\beta} e_{\alpha}^a$
 and $\partial_{\mu}\omega_{\nu}- \partial_{\nu}\omega_{\mu}=
{\displaystyle\varepsilon_{\mu\nu} R\over\displaystyle  4e}$ to obtain
$R= 2(\partial^2\delta_{\mu\nu}-\partial_{\mu}\partial_{\nu}) h_{\nu\mu}(x) $,
up to the
first order terms in $h$. Thus (\ref{eq:an5}) becomes
\begin{equation}
{\delta(\Phi_{+}^{(1)}-\Phi_{-}^{(1)})\over \delta \theta(x)}=
{\Lambda\over|\Lambda|} {1\over 96\pi}R
 \label{eq:an6}
\end{equation}
This agrees with  the well known result for the Lorentz anomaly\cite{kn:H.}.
 In a similar way we can study the response of the overlap to  general
coordinate transformations of 2-dimensional manifold and recover the
result of \cite{kn:H.}for the anomalous divergence of the energy momentum
tensor.

\end{document}